\documentclass[
 twocolumn,amssymb,aps,prb]{revtex4}

\usepackage{graphicx}

\begin{document}

\title{Photovoltaic Oscillations Due to Edge-Magnetoplasmon Modes in a Very-High Mobility 2D Electron Gas}

\author{K. Stone}
\affiliation{Department of Physics and Astronomy, Rice University,
Houston, Texas 77005}
\author{C. L. Yang}
\affiliation{Department of Physics and Astronomy, Rice University,
Houston, Texas 77005}
\author{Z. Q. Yuan}
\affiliation{Department of Physics and Astronomy, Rice University,
Houston, Texas 77005}
\author{R. R. Du}
\affiliation{Department of Physics and Astronomy, Rice University,
Houston, Texas 77005}
\author{L. N. Pfeiffer}
\affiliation{Bell Laboratories, Alcatel-Lucent, Murray Hill, New
Jersey 07974}
\author{K. W. West}
\affiliation{Bell Laboratories, Alcatel-Lucent, Murray Hill, New
Jersey 07974}

\begin{abstract}
Using very-high mobility GaAs/Al$_{x}$Ga$_{1-x}$As 2D electron Hall
bar samples, we have experimentally studied the
photoresistance/photovoltaic oscillations induced by microwave
irradiation in the regime where both 1/$B$ and $B$-periodic
oscillations can be observed.  In the frequency range between 27 and
130 $GHz$ we found that these two types of oscillations are
decoupled from each other, consistent with the respective models
that 1/$B$ oscillations occur in bulk while the $B$-oscillations
occur along the edges of the Hall bars. In contrast to the original
report of this phenomenon (Ref. 1) the periodicity of the
$B$-oscillations in our samples are found to be independent of $L$,
the length of the Hall bar section between voltage measuring leads.
\end{abstract}

%\pacs{1}

%\keywords{Electron transport, Plasmon, Microwave, Semiconductors}

\maketitle

The magnetoplasmons \cite{2} in a two-dimensional electron gas
(2DEG) are coupled modes of 2D plasmons and cyclotron orbits in the
presence of a perpendicular magnetic field, $B$.  It has long been
known that on the edge of the 2DEG, chiral edge modes, or edge
magnetoplasmons (EMP), can also exist \cite{2,3,4,5,6,7,8,9,10,11}.
In transport experiments, both types of mode have been observed in
the microwave photoresistance measurements \cite{1,8,12,13}. One
remarkable recent result is the discovery of a new type of
$B$-periodic oscillation in Hall bar samples consisting of a
high-mobility GaAs/Al$_{x}$Ga$_{1-x}$As heterostructure \cite{7},
where the oscillations are explained by the interference effect
between EMPs being emitted from adjacent electrical contacts and
propagating along the same edge of a Hall bar.  Quantitatively, the
period of the oscillation is found to be $\Delta B$ $\propto$
$n_{s}/\omega L$, where $n_{s}$ is the 2D electron density, $\omega$
$=$ $2\pi f$ the microwave frequency, and $L$ the distance between
the contacts.

\begin{figure}
\includegraphics{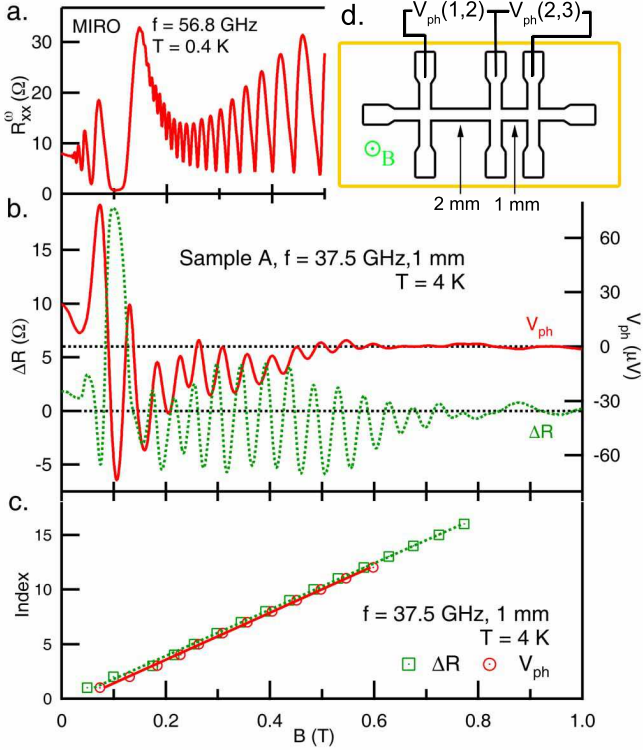}
\caption{\label{fig1} (Color Online) (a) Typical low temperature (T
= 0.4 K) photoresistance data showing microwave-induced resistance
oscillations and the corresponding zero-resistance state; (b)
Differential photoresistance ($\Delta R$) and photovoltage
($V_{ph}$) were measured simultaneously at 37.5 $GHz$ irradiation
along the 1 $mm$ section of sample A, at $T$ $=$ 4 $K$. $B$-periodic
oscillations are clearly observed at $B$ $>$ 0.2 $T$; at lower
magnetic fields, MIRO also contributes to the signal.  (c) Plot of
the field position of oscillation maxima vs. their index. Linear
fits give a slope of 21.16/$T$ and 21.47/$T$ for $\Delta R$ and
$V_{ph}$, respectively. (d) Schematic of Hall bar sample, geometry
of the waveguide cross section, and the contact configuration for
the photovoltage measurement.}
\end{figure}

The $B$-period oscillations have so far been observed \cite{1,8} in
the magnetic field range corresponding to $\omega_{c}$ $>$ $\omega$,
where $\omega_{c}$ $=$ $eB/m^{*}$ is the cyclotron frequency , and
$m^{*}$ $=$ 0.068 $m_{e}$ is the electron effective mass in GaAs. In
a lower $B$ range, $\omega_{c}$ $\leq$ $\omega$, the
microwave-induced resistance oscilations (MIRO), which are periodic
in 1/$B$, are observed \cite{13}. As the sample's mobility
increases, the MIRO become stronger.  It is then possible to create
zero-resistance states (ZRS) under sufficiently high microwave
irradiation \cite{14,15}. While the exact nature of these effects
has not been conclusively established, it has been theoretically
proposed that the MIRO and ZRS result from the non-equilibrium 2DEG
driven by MW irradiation and the consequent symmetry-breaking into
current domains \cite{16,17}.

We have measured the MW photoresistance and photovoltage in a
very-high mobility (mobility $\mu$ $\approx$ $1\times10^{7}
cm^{2}/Vs$) 2DEG in GaAs/Al$_{x}$Ga$_{1-x}$As heterostructures.
Contrast to previous results, in our samples we are able to observe
both the MIRO/ZRS effects and the $B$-periodic oscillations (i.e.,
the EMP modes) in the same sample, where both features overlap at a
certain magnetic field and MW frequency range.  Analysis shows that
these two types of oscillations are essentially decoupled from each
other. Detailed comparisons of data in the $B$-periodic regime
confirm the previous conclusion \cite{1} that the period, $\Delta
B$, is inversely proportional to the MW frequency.  However, in
stark contrast to those reported by Kukushkin \emph{et al.}
\cite{1}, in our very-high mobility samples $\Delta B$ is found to
be independent of $L$, the distance between the contacts. Such
observation is not understood at this point, but it could be
indicative that the photoresponse to the EMP is strongly nonlocal in
our very-high mobility samples.

Our specimens are Hall bars defined by lithography and wet-etching
from a GaAs/Al$_{0.3}$Ga$_{0.7}$As heterostructure grown by
molecular-beam-epitaxy.  Sample A has a Hall bar width ($W$) of 100
$\mu m$ and has two sections of 1 and 2 $mm$ ($L$) that make up 10
and 20 square sections respectively.  The contact leads have the
same width of 100 $\mu m$.  Sample B has a similar geometry but with
a width (both bar and leads) of 200 $\mu m$, giving the 1 and 2 $mm$
sections 5 and 10 squares respectively.  High quality Ohmic contacts
to the 2DEG were made by high temperature diffusion of Indium. After
a brief illumination with visible light, at $T$ = 0.3 $K$ the sample
A (B) attained a sheet density $n_{s}$ $\approx$ $2.27\times10^{11}
/cm^{2}$ ($2.45\times10^{11} /cm^{2}$), and a mobility $\mu$
$\approx$ $8.3\times10^{6} cm^{2}/Vs$ ($11\times10^{6} cm^{2}/Vs$).
Our measurement was performed in a $^{3}He$-refrigerator equipped
with a superconducting magnet. The microwaves (MW) were generated by
a set of Gunn diodes and sent via a rectangular waveguide (WG-28) to
the sample immersed in the $^{3}He$ coolant. The mutual orientations
of the waveguide, 2DEG plane, and the magnetic field corresponded to
Faraday configuration. For MW frequencies $f$ $<$ 44 $GHz$, the
waveguide operated in single-mode and the $\vec{E}$ polarization of
the MW was perpendicular to the Hall bar direction.

The magnetoresistance under MW irradiation, $R^{\omega}_{xx}$, was
measured by standard low frequency (23 Hz) lock-in technique and
using a sample excitation current $I$ = 1 $\mu A$. The
photoresistance, $\Delta R$ = $R^{\omega}_{xx} - R^{0}_{xx}$, where
$R^{0}_{xx}$ is the ``dark" resistance, was measured by a
double-modulation technique, whereas $V_{ph}$, the photovoltage
signal, was measured in the absence of an external excitation
current $I$.

Figure 1 shows typical data of $R^{\omega}_{xx}$, $\Delta R$, and
$V_{ph}$ measured in sample A.  We first notice that in the low $B$
range, $B$ $<$ 0.2 $T$, and at a low temperature $T$ $=$ 0.4 $K$,
the $R^{\omega}_{xx}$ is completely dominated by the MIRO signal
(FIG. 1a), and in an increasing $B$ range, $B$ $>$ 0.2 $T$, usual
Shubinikov de Haas (SdH) oscillations take over.  Through the
contacts for the 10-square Hall bar section in sample A, strong
$B$-periodic oscillations in both $\Delta R$ and $V_{ph}$ are
observed (shown, for example, in FIG. 1b for $\omega$ = 38.9 $GHz$)
in a wide temperature range from 0.4 to 20 $K$. These are the
characteristic signals that originate from the EMP, as first
reported by Kukushkin \emph{et al} \cite{1}. Conforming to their
results, the period of the oscillations, $\Delta B$, are found to be
identical in $\Delta R$ and $V_{ph}$ (FIG. 1c) with some phase shift
with respect to each other.

The $\Delta R$ (38.9 $GHz$) for sample A at various $T$ is shown in
FIG. 2.  At $T$ $=$ 0.3 $K$ and in the range $B$ $<$ 0.2 $T$, sharp
MIRO can be identified as a pair of peaks and valleys around the
$\omega/\omega_{c}$ $=$ 1(solid arrow).  Beyond $\omega/\omega_{c}$
$=$ 1, a second pair of peaks and valleys surrounding
$\omega/\omega_{c}$ $=$ 1/2 (dotted arrow) is also observed; this is
the photoresistance feature associated with the two-photon nonlinear
process reported elsewhere \cite{13}.  In an increasing $B$ range,
$B$ $>$ 0.2 $T$, the $\Delta R$ is dominated by 1/$B$ oscillations
originating from the MW heating effect in SdH; the oscillations are
180$^{\circ}$ out of phase with respect to the SdH in $R_{xx}$.  In
general, the $B$-period oscillations are best resolved at an
elevated temperature ($T$ $>$ 4 $K$), where both MIRO and SdH are
damped out.  In our samples, the $B$-periodic oscillations are found
to persist up to 20 $K$.  This temperature is somewhat lower than
that reported in Kukushkin \emph{et al.}; the difference could be
due to the fact that we use a relatively low MW power (typically
10-100 $\mu W$ at the sample location).

\begin{figure}
\includegraphics{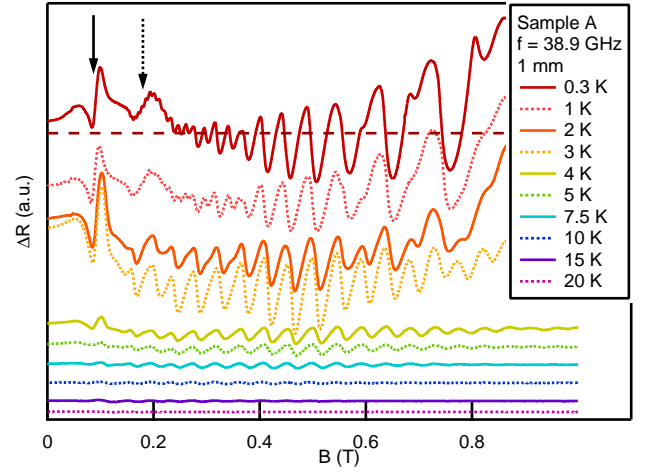}
\caption{\label{fig2}  (Color Online) Temperature dependence of the
photoresistance, $\Delta R$, for sample A at a microwave frequency
38.9 $GHz$. The arrows mark the microwave-induced resistance
oscillations (MIRO). At low temperature, T $<$ 2 $K$, the $\Delta R$
is dominated by 1/$B$ oscillation (MIRO and SdH) and at higher
temperatures, T $>$ 4 $K$, by $B$-periodic oscillations.}
\end{figure}

We have carefully measured $V_{ph}$ for samples A and B over a
frequency range from 27 $GHz$ to 130 $GHz$ and found that
$B$-periodic oscillations are generic features in our samples. As
examples, the $V_{ph}$ for the sample A (1 $mm$ section) is shown in
FIG. 3a for various values of $f$ linear fits to the $B$-positions
of the peaks vs. their index (FIG. 3b) are generally satisfactory.
As shown in Fig. 4c, our data in particular confirms the reported
\cite{1} inverse-linear relation $\Delta B$ $\propto$ $1/\omega$.
4c.

\begin{figure}
\includegraphics{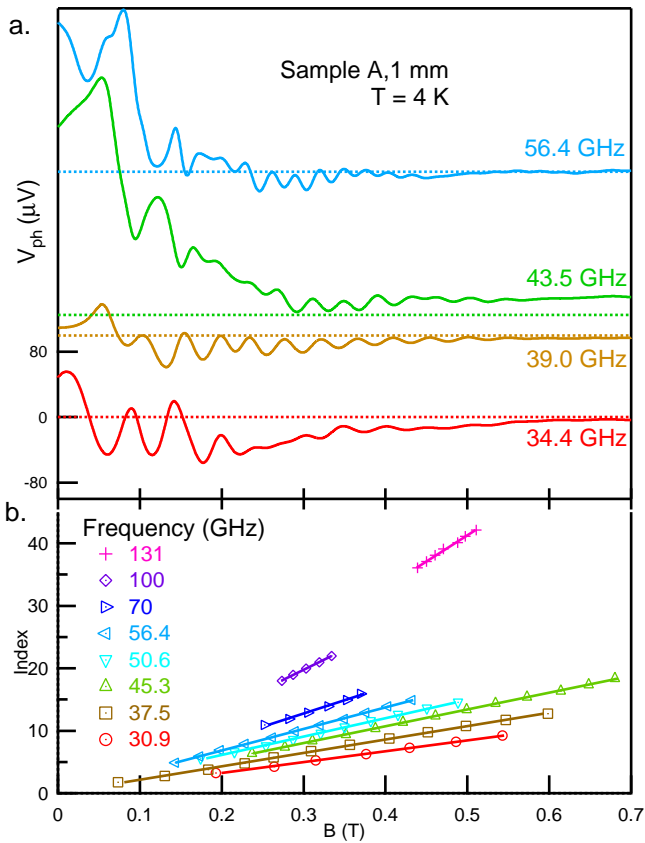}
\caption{\label{fig3} (Color Online) (a) Examples for photovoltage
signal, $V_{ph}$, as a function of the magnetic field, $B$,  for
selected microwave frequencies. (b) Linear fits to $B$ positions of
the $V_{ph}$ peaks show that the oscillations are periodic in $B$;
from the slope of the fitting line the oscillation period, $\Delta
B$, is determined. }
\end{figure}

While the data from our very-high mobility samples have by and large
confirmed the reported $B$-periodic oscillations and its inverse
scaling with the MW frequency, we found that the period, $\Delta B$,
appears to be independent of $L$ for both samples A and B.  Such a
lack of length scaling can be clearly shown in the data presented in
FIG. 4. Taking sample A and referring to FIG. 1d for the contact
configuration, we observe that the $V_{ph}(1, 2)$ and the
$V_{ph}(2,3)$ have almost identical oscillations except that the
phase differs by $180^{\circ}$.  Since the leads were both connected
the same way and the center junction of the Hall bar lines up
closely with the center of the waveguide, we can assume that the
opposite signs in $V_{ph}$ originate from the opposite gradients of
the microwave electric field inside the waveguide.  The fact that
the contacts for the 1 $mm$ and 2 $mm$ sections have resulted in a
``mirror image" $V_{ph}$ was completely unexpected. We have measured
the same quantities over a wide range of frequencies, and found
consistent results.  In particular, we found that the relationship
between the period $\Delta B$ and $1/f$ is almost identical for the
1 $mm$ and 2 $mm$ portion of both samples with slopes 1.86 $T\cdot
GHz$ and 1.84 $T\cdot GHz$, respectively. From these results we
conclude that the $B$-periodic oscillations are not visibly
dependent on length of the Hall bar between the leads. To confirm
this result, we also measured $V_{ph}$ in sample B which had a width
of 200 $\mu m$ and respective sections are 5 and 10 squares. As
shown in FIG. 3b, the $B$-periodic oscillations in this second
sample also have no $L$ dependence.

\begin{figure}
\includegraphics{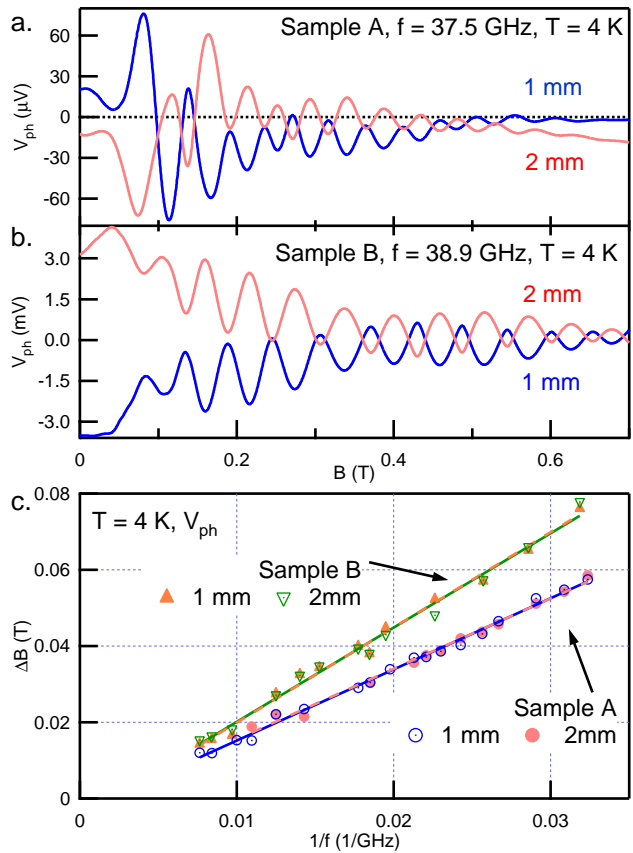}
\caption{\label{fig4} (Color Online) (a) Example traces of
photovoltages $V_{ph} (1,2)$ (measured in the 2 $mm$ section) and
$V_{ph} (2,3)$ (1 $mm$ section) in sample A, at a microwave
frequency 37.5 $GHz$. The $V_{ph}$ oscillations from the 2 sections
have approximately the same magnitude but with opposite signs. (b)
Examples of similar data from sample B at 38.9 $GHz$. (c) The plot
of $\Delta B$ vs. inverse microwave frequency 1/$f$ shows a linear
relation; the data from 1 $mm$ and 2 $mm$ sections from the same
sample collapse on the same line, indicating that the $\Delta B$ is
independent of contacts in a wide frequency range. From the plot of
$\Delta B$ vs 1/$f$, slopes of 1.85 $T \cdot GHz$ and 2.50 $T \cdot
GHz$ are determined for samples A and B, respectively.}
\end{figure}

Since samples A and B have different electron densities $n_{s}$, it
is worthwhile to examine if $\Delta B$ scales with $n_{s}$. From
FIG. 4c, we found a slope ratio of 2.50/1.85 $=$ 1.35, which is
about 25$\%$ larger than the ratio of density 2.45/2.27 $=$ 1.08. We
consider this to be qualitatively consistent with the result from
Kukushkin \emph{et al.}, i.e., $\Delta B$ $\propto$ $n_{s}$.

To summarize, we have experimentally measured the photoresistance
and photvoltage signals in a very-high mobility 2DEG, and observed
$B$-periodic oscillations over a wide range of microwave
frequencies.  Both the 1/$B$ oscillations which originate from
electron Landau level transitions, and the $B$-periodic oscillations
which originate from edge magneto-plasmon modes, have been observed
in the same sample.  We confirm that the oscillation period of the
later inversely scales with the microwave frequency, $\Delta B$
$\propto$ 1/$\omega$.  However, our data in the very-high mobility
samples do not conform to the previous result \cite{1} that $\Delta
B$ $\propto$ 1/$L$.

Our observations from the very-high mobility samples are rather
puzzling and we cannot find a satisfactory explanation at this
point.  On one hand, several major features observed in our samples
clearly support the interpretation of the $B$-periodic oscillations
as being associated with the EMP. In particular, $\Delta B$ is
proportional to $n_{s}/\omega$, showing a characteristic relation
for propagating EMP modes in a 2DEG.  On the other hand, without a
length scale, the interpretation based on a simplistic interference
model \cite{1} appears irrelevant.  Even more puzzling is the fact
that both the 1 $mm$ and 2 $mm$ sections exhibit photovoltage
signals that completely mirror each other.

Our results strongly suggest that in very-high mobility Hall samples
the microwave photoresponse are predominately nonlocal, in the sense
that the photovoltaic signal can propagate along the edge of the
entire sample.  It is plausible that long-wavelength EMP modes,
which are chiral in nature, can circulate around the whole sample
perimeter and dominate the dynamics of the 2DEG. This process would
be more favorable in our very-high mobility samples due to the
extremely long decay length of the EMP modes in these samples.
Moreover, the photocurrents that are presumably generated by
dragging of the EMP propagation could be strongly dependent on the
EMP wavevector (both in terms of magnitude and direction).
Consequently, the interference signal generated in one pair of
contacts could propagate along the edge of the entire sample and
dominate the photovoltaic signal on the other pairs.  In an
alternative scenario, the mirror $V_{ph}$ signals measured from the
adjacent contact pairs could be driven by the Hall field in the
presence of photocurrents running in the central contact lead,
exciting oscillations with the same period in $B$.

In conclusion, we have experimentally investigated the
magneto-oscillations in both resistance and photovoltage in a
very-high mobility 2D electron gas irradiated by $GHz$ microwaves.
For the first time, MIRO, ZRS, and EMP were observed in the same
sample, indicating that these are decoupled effects. Our central
finding, that the $B$-periodic oscillations observed in our samples
have no apparent dependence on the contact configuration (including
the contact seperation distace) can be interpreted tentatively as
resulting from the chiral dynamics of the EMP and the strong
nonlinear response in photovoltage to such properties.

%\begin{acknowledgments}
The research at Rice is supported by the NSF grants DMR-0408671 and
DMR-0700478.  Use of the Rice Shared Equipment Authority (SEA)
facilities is appreciated.
%\end{acknowledgments}

\end{document}